# Rating models: emerging market distinctions[*]

ALEXANDER KARMINSKY[†]

Abstract

The Basel II Accords have sparked increased interest in the development of approaches based on internal ratings systems and have initiated the elaboration of models for remote ratings forecasts based on external ones as part of Risk Management and Early Warning Systems. This article evaluates the peculiarities of current ratings systems and addresses specific issues of development of econometrical rating models for emerging market companies. Financial indicators, market-value appraisals and macroeconomic indicators of different countries were used as explanatory variables. Standard & Poor's and Moody's ratings were considered as modeling ratings.

*Keywords*: corporate ratings, models, financial risk.

*JEL codes*: G21, G32.

[*] Series WP7 «Mathematical methods for decision making in economics, business and politics»; Paper WP7/2010/06; Original text:
https://www.hse.ru/data/2010/09/23/1223926390/WP7_2010_06.pdf
[†] † State University Higher School of Economics, 109028, Moscow, Russia, Pokrovskiy bul., 11, J-417; e-mail: akarminsky@hse.ru). I would like to thank Anatoly Peresetsky for his helpful comments.

# 1. Introduction

Ratings are in high demand in market-driven economies. Within a business setting, the rating process has a moral component. A Rating agency's reputational capital serves as a regulatory element (Partnoy, 2002). In addition to independent appraisals of investment risk in the form of the rating agency's opinion, ratings also function as a sort of licensing.

The Basel II Accord (Basel, 2004) has sparked increased interest in ratings and their models. The development of approaches based on internal ratings systems has a practical interest, especially for developing markets. The topic has received increased attention in connection with the global crisis that began in 2007.

In this work, we analyzed possibilities for modeling ratings applied to industrial companies and banks of developing countries. Emphasis was placed on the elaboration of econometric models. As explanatory variables, financial indicators (which characterize the activities of a company), market indicators (which reflect the dynamics of its stock quotations), and macroeconomic variables and dummies of industrial and country affiliations were employed.

Ratings of Moody's Investors Service and Standard & Poor's agencies were considered as modeling ratings. This made the evaluation of the specific approaches of each of these agencies possible. Samples were made up of data from these agencies and the Bloomberg information agency.

Analysis of the predictive power of the derived econometric models allowed for an appraisal of these models to be made. Particular attention was paid to variables in rating models in accordance with their affiliations with developing countries or with particular industries.

It was shown that industry and group of countries affiliation influences ratings.

The work consists of seven sections. The second section addresses the particularities of ratings as a measurement of risk in Russia and the countries of Central and Eastern Europe. A comparison of the methodologies of the two leading agencies was the subject of the third section. In Section 4, there is an examination of the types of models used and the formation and statistical characteristics of the samples.

Models of corporate ratings and bank ratings pertaining to developing markets and a comparison of ratings of Moody's and S&P were systematized in Sections 5 and 6. Conclusions were presented in the final section.

## 2. Development of ratings services in developing countries

To begin with, we will consider rating opportunities in Russia. We can observe several waves of interest in these instruments. The entry of the international rating agencies and the ratings they made in Russia (beginning in 1996) had little impact before the 1998 financial crisis or immediately after it.

The opportunity for foreign borrowing, including borrowing by industrial companies beginning in 2003, gave impetus to their development. The number of ratable objects has more than tripled since then, reaching more than 300 at the beginning of 2009 (about half of them are banks and more than a third are companies.) The process was encouraged when Russia received investment-level ratings in 2005-2006.

The crisis of 2008-2009 has had an effect on the rating process. A number of ratings were withdrawn. Russia's sovereign ratings were lowered by Standard & Poor's and Fitch Ratings by one grade, although the ratings remained on the investment level at BBB. The insignificant lowering of sovereign ratings did not dampen interest in them from economically active objects, as happened in 1998.

Although a large portion of bank ratings was assigned by Moody's Investors Service (hereinafter, "Moody's"), the Standard & Poor's agency (hereinafter "S&P") leads in ratings of industrial companies (Fig. 1) and their financial instruments.

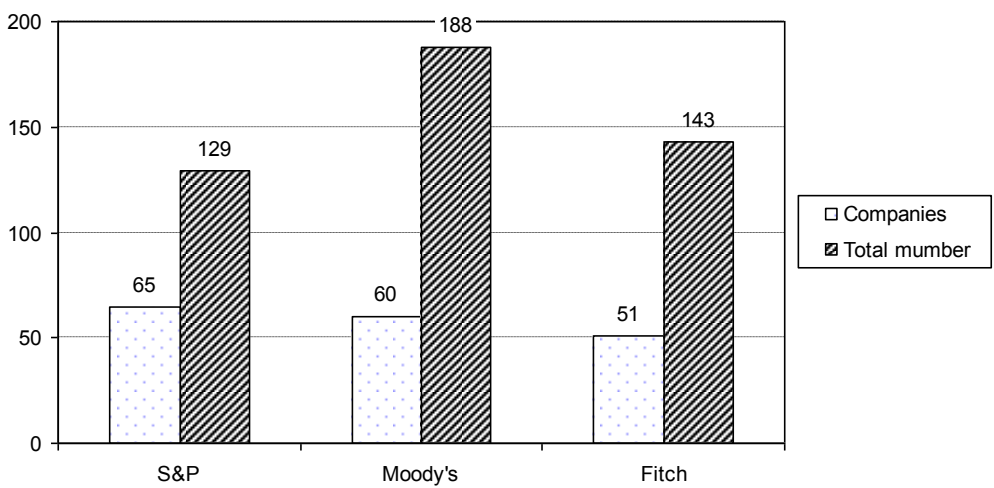

**Fig. 1.** Distribution of the total number of ratings and ratings of industrial companies by international agencies (January 2009)

The distribution of international agencies' corporate ratings by grades (Fig. 2) shows that the level of ratings of Russian companies is comparatively low. Less than 20 companies have investment-level ratings.

The average level of ratings was between BB- and BB for all three agencies, while the average level for S&P was almost BB- and for Moody's it was Ba2 which is equivalent to BB. For the Fitch Ratings agency, the average level was between those grades.

The average rating level of companies is higher than that of banks. Although S&P ratings were practically identical for banks and companies alike, companies and regions ratings by Fitch were substantially lower for banks (by more than 0.5 and 1.5 grades on average respectively). This applies even to a greater degree to Moody's. This distinction is not only methodological, but also connected with the Moody's wider scope of bank ratings.

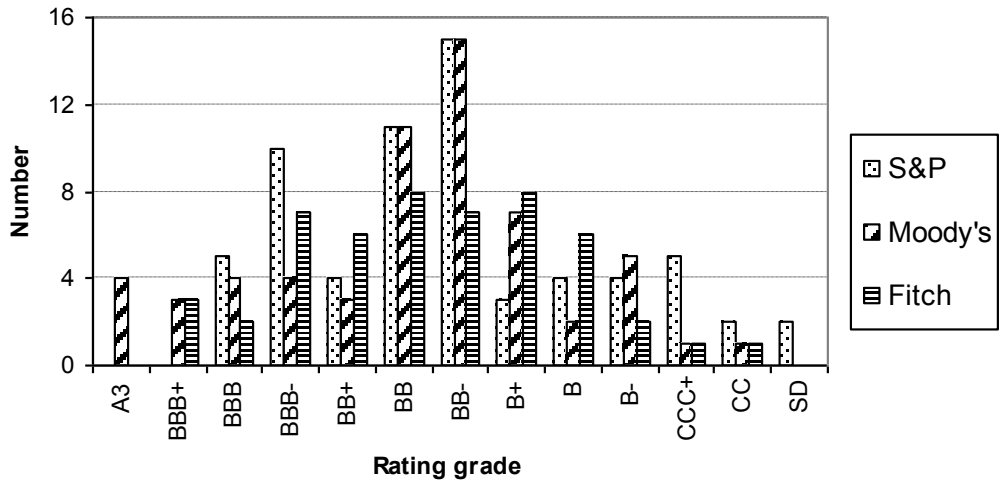

**Fig. 2.** Distribution of ratings of Russian companies by grades (January 2009)

The ratings of banks and industrial companies in Central and Eastern Europe (CEE) have much in common because they are related to developing countries. These countries' development level lags considerably behind that of the original members of the European Union. However, these countries were oriented toward membership in the EU from the very beginning, and many of them have now become members.

Data analysis shows that sovereign ratings of the countries of the EU are in the lower part of the investment range (A and lower), with the exception of the higher ratings of Slovakia, Czech Republic (A+) and especially Slovenia (AA). Serbia, Macedonia and Bosnia and Herzegovina have speculative ratings.

At the same time, companies and banks have mainly speculative level ratings and the number of rated objects was low (statistical data on the comparison between country ratings and banks ratings is presented in Appendix 1). This is largely explained by European Union's support for these countries, although this support has been limited since the onset of the economical crisis. There are few large companies in these countries. The ones that do exist are subsidiaries of transnational companies, and this may lower the ratings level, considering that their market capitalization (or volume of assets) is one of the main explanatory variables for these ratings.

Thus, despite higher sovereign ratings, company ratings in these countries are on the same level as Russian companies: mainly in the upper range of speculative ratings.

Despite the comparative growth in the number of ratings, there are still clearly too few of them in developing countries. In addition, rating methods are largely inexplicit, and expertise plays a significant role in them. This hinders the use of ratings for risk evaluation and decision making even on the state level. This is the reason for the interest in the creation of internal ratings and model ratings that can be used as preliminary evaluations in making management decisions.

# 3. Overview of the literature and methodology
## 3.1. Particulars of corporate credit rating methodology

The credit ratings of borrowing companies are composed of two elements: a business analysis and an analysis of the financial profile (S&P, 2009). Besides an analysis of financial indicators, the procedure for assigning a rating consists of research of the fundamental characteristics of the business such as country risks, the structure of the industry and its perspectives for growth, the company's competitive advantage, the system of regulation, as well as management and strategy. Many of these factors are qualitative, although the results of future activity depend on them. These factors are also more significant for the speculative level of ratings.

Each of the elements in the process of preparing a ratings statement is subjected to a complex examination. The specifics of the industry and any nonstandard operations are taken into account during a financial anaylisis. The company's financial policy and its approach to risk management are analyzed. Attention is mainly given to cash flow, coefficients of debt and interest settlement with cash flow and funds from operations and liquidity. The checklist of basic financial indicators influenced the choice of the set of explanatory variables in the models discussed below.

Moody's agency also gives financial indicators a significant role in the assignment of its ratings, in addition to analysis of country and industry factors. The agency uses 11 key financial indicators (Moody's, 2009), a number of which are strongly correlated and have exerted influence on the set of explanatory variables in the models.

With regards to the banking sector, factors of support have particular significance, and have been methodologically incorporated in (Moody's, 2007). A two-level system of ratings has been formed, in which financial stability ratings (Bank Financial Strength Ratings) determine the level of creditworthiness of financial institutions without support (stand alone), whereas Deposit Ratings take support into account.  Similar steps have been taken recently by the S&P, which also considers factors of support from the state and from parent companies.

Our long-term goal is to research the possibility of forecasting bank and company ratings based solely on publicly accessible information, including indicators from international financial reports and market conditions on stock exchanges.

## 3.2. Overview of literature on modeling corporate ratings

Changes in ratings play an important role in transactions with interest-rate risks. Despite a decrease in the normative significance of ratings, their presence and popularity have grown since the mid-1970s (Partnoy, 2002). This is due largely to the regulatory significance of ratings, in addition to their market significance (Karminsky, Peresetsky, 2009).

Although the initial application of ratings was to debt obligations, corporate ratings are now steadily gaining significance (Altman and Suggitt, 2000) for the organization of syndicated credits, the rating of corporate bonds and other purposes (Altman and Saunders, 1998; Servigny and Renault, 2004; Partnoy, 2002).

A number of researchers have shown (Altman and Rijken, 2004; Pederzoli and Torricelli, 2005; Curry et al., 2008) that ratings are cyclical. There is a certain lag between the recording of financial results and the changing of a rating. While this creates stability in the rating process and averts reactions to comparatively insignificant events, it does not always provide for a timely reaction by ratings agencies to significant procedures. The widely-discussed collapse of several major companies in the last several years is an example of this (Servigny and Renault, 2004).

At the same time, as (Amato and Furfine, 2004) showed that credit ratings have risen less during times of recession using U.S. firms and the S&P data. However, ratings do not display excessive sensitivity to business cycles. The potential time degradation of ratings (which is due to a change in the extent of credit risk) deserves attention. Without discounting this factor, especially for new financial instruments (at present predominantly structured transactions), it should be noted that degradation may be connected with the dynamics of the market as a whole. In regards to banks, the absence of such degradation during the transition to ordinal scales was shown in (Karminsky, Peresetsky, 2007).

The reason for the relative volatility of ratings is the specifics of the assignment of country ratings, especially for developing countries (Moody's, 2007). In (Kaminsky and Schmucler, 2002; Reinhart, 2002), it is shown that there are three possible channels of instability that arise from changes to a country rating during a crisis:

- directly through the value of debt obligations and stock on the market
- through contagion and generated global unstability
- due to markets in countries with lower ratings because of their greater liability to fluctuations

In (Reinhart, 2002), it is also shown that changes in sovereign ratings influence the spread and income of bonds, complicate access to resources on developing markets, hasten the transition from currency crisis to banking crisis and may impact recession. Some of the specifics of the rating process during crisis, including in regards to developing countries, are also examined in (Joo and Pruitt, 2006). Significant attention has been given to the analysis of indicators of financial and banking crises. Developing countries have been a mainstay of this research (Kaminsky and Schmucler, 2002; Rojas-Suarez, 2002).

The evaluation of sufficiency of capital as a measurement of risk based on internal ratings according the IRB Approach of Basel II, may be based on probability of default or rating models (Basel, 2004). It may also use an evaluation of transition matrices and the mechanism of Markov chains

(Frydman and Schuermann, 2008) or econometric models, including scoring (Altman and Saunders, 1998; Altman, 2005; Feng et al., 2008).

A number of works have been devoted to the elaboration of internal ratings systems and early warning. An overview of methodological specifics of elaborating models is made in (Altman and Saunders, 1998; Karminsky et al., 2005).

In (Carey and Hrycay, 2001), specifics of the joint use of several methods for the evaluation of the probability of default on debt instruments according to an internal rating scale were examined. Mapping to a standardized scale and scoring models were implemented. The presence of a data series of long duration is critical. A number of the specifics of the elaboration of internal ratings systems are also examined in (Jacobson et al., 2006; Servigny and Renault, 2004; Hanson and Schuermann, 2006). In the last of these works, the confidence interval technique was used to refine rating gradation.

Selection of the explanatory variables is methodologically important for the elaboration of corporate ratings models. The indicators that are employed by the rating agencies may be used (Moody's, 2009; S&P, 2008). Others which have been employed by researchers may be used (Rojas-Suarez, 2002; Servigny and Renault, 2004; Guttler and Wahrenburg, 2007; Curry et al., 2008). Typical indicators are the size of the company, its profitability, stability, liquidity and structure of the business, as expressed through companies' balance-sheet figures. In recent years, the use of such factors as state support for companies, and support from the parent company or group of companies has become more prominent (Moody's, 2007; S&P, 2009).

The use of macroeconomic indicators has also become more typical recently (Carling et al., 2007; Curry et al., 2008; Peresetsky, Karminsky, 2008). Among the most common indicators are inflation index, real GDP growth, industrial production growth and, for export-oriented countries, oil prices and changes in the cross-rate of currencies. Separate mention should be made of market indicators (Curry et al., 2008), which is especially important for publicly held companies (market value of companies, volatility of stock prices, systemic parameters, etc.).

It should also be noted that alternate indicators can also be used for developing countries (Altman, 2005; Rojas-Suarez, 2002; Karminsky et al., 2005). These are characteristic of developing markets and predominantly speculative ratings. They include the value of resources, percent margin, pace of asset growth and growth of interbank debt (also on an international level).

Variation over time, both of the dimensions of the risk and the approaches of the rating agencies, points to the use of a time factor in models. The use of panel data may also be incorporated. Some of the specifics of these approaches are found in (Elton et al., 2004; Frydman and Schuermann, 2008).

The particular significance of industry affiliation and possible differences among ratings of companies of varied profiles and regions can be noted for corporate ratings (Niemann et al., 2008). This is connected with the specifics of business in various segments of production activity. Industry-specific models and the use of dummies depending on the industry and country of companies are possible.

A number of works have noted differences in the ratings of various agencies (Packer, 2002; Bae, Klein, 1997; Kish et al., 1999). Corresponding factors of national and international agencies were analyzed in those works. In practically all the research, the two main rating agencies, Moody's and the S & P, were considered.

3.3. Ratings during the global financial crisis: is there an outlook for growth?

The global financial crisis exposed a number of problems of the ratings business and the entire financial management system. The financial system grew markedly in the first decade of the 2000s (IMF, 2009). New financial instruments were created. They were designed for higher profit with lower expected levels of risk. However, the inability of regulatory organs and rating agencies to evaluate the threat from the asset price bubble appeared limited.

Faith in light regulation based on the discipline of the financial market and hope for the successful distribution of risks through financial innovation does not preclude their concentration. An IMF analysis shows problems on three levels:

- Financial regulatory and monitoring organs proved to be incapable of exposing a higher concentration of risk brought on by rapid growth in financial innovation.
- No account was taken of growing macroeconomic imbalances that contributed to the growth of systemic risk in the financial system and real estate market.
- International financial organizations and the monitoring and control system that was in place were unable to cooperate reliably on the international level to identify vulnerable areas in transnational relations.

Heightened possibilities of infection during the liquidity deficit are noted in (Karas et al., 2008) in regard to the Russian banking stystem and developing markets. It is shown there that regulating the liquidity of individual banks is not sufficient to avoid a systemic crisis. Resource management by the lender of last resort is necessary to restore the coordination of the MBK.

The crisis emphasized the need for clearer signals in economic policy and the development of international cooperation on a number of economic and financial issues including ratings. Among the measures suggested was taking leadership in responsive measures to systemic global risk. The establishment of an early warning system is a good example of this.

Development and regulation of early warning systems requires improved independent evaluation. Rating agencies should play a crucial role in this process. Policy has to be coordinated in various areas, including supervision of the rating agencies, bookkeeping practices and auditing. These measures should be coordinated both within a country and on the international level (IMF, 2009).

Among the problems that arise in connection with the financial crisis, rating agencies' lag in the methodology of assigning ratings may be cited. Another factor is the calculation of systemic risks of the global financial system and a lag in the evaluation of complex financial instruments. The more active role agencies have taken in the development of methodology, including areas emphasized by the Basel committee may be noted (Basel, 2009).

## 4. Data and models

### 4.1. Models and rating scales

Multiple choice probit models were used for follow-up studies. Earlier, they had been used for the elaboration of bank rating models (Magnus et al., 2007). Further, three numerical scales were used. These correspond to the use of classes and gradations as scales of ratings as well as a mixed scale which enables the limitation of the sample volume. The mapping of these scales to the numerical scales has 8, 18 and 12 levels, respectively, and a higher rating corresponds to a lower number.

### 4.2. The Sample: financial and market variables, macrovariables and ratings

To elaborate ratings models for industrial enterprises, a sample was made on the principle of affiliation with companies in a number of industries (oil and gas, metals, retail trade, energy, telecommunications and heavy industry) that are potential competitors of Russian production companies of the same profile.

Conditions for selection in the sample were: the presence of an S&P rating at the time the sample was made, affiliation with a selected industry, accessibility of financial data and market indicators, and tradability, as indicated by a liquid market for the company's stock. In the sample, there were 215 companies from 39 countries with S&P ratings as of Spring, 2008.

The sovereign and corporate credit ratings of companies were taken from the websites of the S&P and Moody's agencies as they appeared in March, 2008 (S&P only) and February, 2009. Financial and market indicators were taken from the Bloomberg information system. A minimum of one financial indicator was assumed in each of the following groups: market valuation, size, profitability, market risks, balance-sheet and cash flow.

Distribution of the companies represented in the sample by rating gradation as of Spring, 2008 is presented in Fig. 3. The low number of companies with ratings in categories AAA, AA and B justified the use of a mixed scale with differentiation of gradation in classes A to B.

More than half the companies were represented by five countries: U.S. (74), Russia (31), Canada (15), Great Britain (13) and Japan (10). Division of the countries into developed or developing economies was done according to the methodology of the International Monetary Fund (IMF). The sample contains a greater number of companies from developed countries (152) than from developing (63). Among developing countries additionally to Russia there was Bulgaria, Czech Republic, Hun gary, Poland and other coutries' companies. Among developed countries additionally to USA, Canada, Great Britain and Japan there was practi cally all countries from «old» European Union.

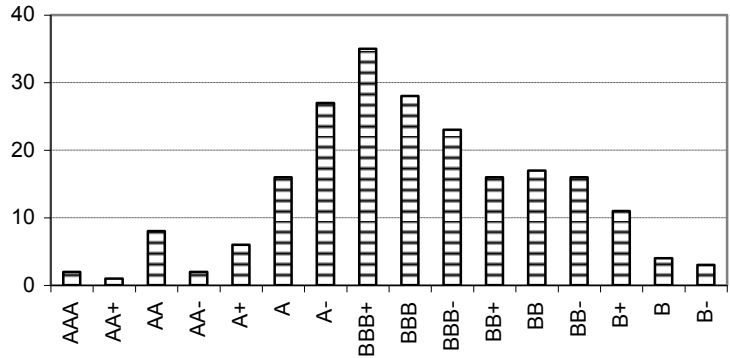

**Fig. 3.** Distribution of companies in the sample by gradation of rating

The time lag between financial indicators and ratings was determined to be 1.5 years (financial indicators were taken from Autumn, 2007 and ratings from the beginning of 2009). This agrees with the conclusions reached in (Karminsky, Peresetsky, 2007) and with a number of other works such as (Altman, Rijken, 2004). Financial and market data for only one accounting period can be considered a flaw in the sample[‡].

The sample for banks included (Peresetsky, Karminsky, 2008) about 1000 observations in 2002-2005 of 380 banks in 42 developing and developed countries. The distribution of the sample by gradation and region, and the statistical distinctions of the sample can be found in this work also.

4.3. Explanatory variables and descriptive statistics

---

[‡] Semiannual lag calculations resulted somewhat lower outcomes. Bank models testify about very flat lag extremum within half to two-year range. The similar hypothesis for companies will be verified to the end of 2010 when the crisis impact on ratings could be assessed. Measures undertaken in the beginning of 2009 to overcome the crisis virtually did not affect both final financial results and ratings.

The list of the main financial and market indicators used for the elaboration of the rating models, as well as their descriptions and the expected influence on the ratings of industrial companies, is presented in Table 1. Descriptive statistics and the regression correlation matrix are contained in Appendix 2. Supplemental indicators have been used in a number of models, as described in the text.

Table 1

Main financial and market indicators used in rating models

| Indicator | Designation | Formula | Group | Expected influence |
|---|---|---|---|---|
| Market capitalization | Market Capitaliza-tion | Stock price* Number of shares (mln USD) | Size | + |
| Return on a share (reciprocal value) | Price/ Earnings | Price per share/Net earnings from one share | Market indicators | – |
| Return on assets | ROA | Net earnings/Average assets | Profitabi-lity | + |
| Operating margin | Operating Margin | Operating revenue/ Receipts | Profitabi-lity | + |
| Volume of borrowing to EBITDA Ratio | Debt/ EBITDA | Debt/Earnings before deductions (interest expenses, taxes and amortization) | Balance-sheet and cash flow (BCF) | + |
| Cash flow per year to receipts Ratio | Cash flow/ Sales | Cash flow/Receipts | Profitabi-lity | + |
| Volume of borrowing to assets Ratio | Debt/Assets | Debt/Assets | BCF | – |
| Long-term debt to capital Ratio | LT Debt/ Total Capital | Long-term debt/Capital | BCF | – |
| Gross profit to interest expenses Ratio | EBITDA/ Interest expenses | Profit before deductions/Interest expenses | Profitabi-lity | + |
| Proxy for current liquidity | Current ratio | Short-term assets/ Short-term liabilities | Liquidity | – |
| Share of fixed assets in assets | Fix Assets/ Total Assets | Fixed assets/Assets | BCF | – |

| Share value to cash flow Ratio | Price/ Cash flow | Share value/Cash flow | Market indicators | + |
| Systemic risk for the last 2 years | Beta | Cov(Ri, Rm)/Var(Rm) | Market indicators | – |
| Volatility of share price in a year | Volatility | Var(Ri)^0.40 | Market indicators | – |

We will characterize market risks by the last three indicators. The Beta coefficient shows how much dividend yield correlates with market yield, and the Volatility indicator characterizes the variability of share price per year.

Country distinctions were characterized by macrovariables: annual rate of inflation; real GDP growth; CPI (Corruption Perception Index); and sovereign rating, which may be seen as a proxy for the institutional environment in which a company functions. The first two indicators were taken from data from 2007 (World Bank), the corruption index comes from Transparency International, and the sovereign ratings were taken from rating agency data. The expected influences of inflation and corruption were negative. The remaining indicators were positive. Higher CPI designations correspond to lower levels of corruption.

A number of dummies were also used in the models. The relationships of companies to countries with developed economies Dev (1- developed, 0- developing) and to Russia were introduced for an analysis of the influence of affiliation to the groups on the rating level. Companies' liability to risks in dependence of its affiliation with various industries was traced through the introduction of a dummy of affiliation in the following sectors: telecommunication, oil and gas, metal and mining, consumer, utilities, and manufacturing and chemicals.

The choice of explanatory variables for banks was created the same way (Karminsky, Peresetsky, 2007; Peresetsky, Karminsky, 2008). Variables used in the models will be commented on in the analysis of the resulting tables.

## 5. Econometric corporate rating models
### 5.1. Base model and its improvement

The models examined in this work depend exclusively on open information. Accordingly, we examined possibilities derived from the use of indicators based on company financial accounts prepared to international standards and supplemental possibilities provided by macrovariables and market elements.

Among the questions that we faced in elaborating rating models, we emphasized the following:

- Do the ratings of enterprises depend on their affiliation with a group of countries (developing countries, Russia)?
- Do ratings depend on affiliation with an industry?
- Is it possible to incorporate a high enough level of information in sovereign ratings using macrovariables?

In the elaboration of the base rating model for the Standard & Poor's agency, indicators were chosen from each group of financial indicators. We considered the capitalization of the company as the indicator of its size for all models. As criteria for comparison at the first stage, statistical characteristics of the quality of the models (Pseudo-$R^2$, t-statistics) were used, to which predictive characteristics were added at the next stage.

The models derived with the use of the scale for rating classes as a dependent variable were presented in Table 2. Coefficient signs matched prior expectations.

Positive influence on the rating level is exerted by factors such as a company's market capitalization, return on assets and level of income in relation to interest expenses. Positive influence on the operating margin was expected. It is the basis of a company's stability.

The negative influence of the relationship of long-term debt to capital is also intuitive, since rating agencies closely follow the level of borrowing and the probability of repayment. The negative influence of the proxy for current liquidity is natural, since it is the reciprocal value of the current liquidity indicator. Cross-impact of financial indicators was not revealed.

Table 2

Rating class models

|  | Model number and agency (sp = S&P; mo = Moody's) | | | |
|---|---|---|---|---|
|  | Base S&P | Quadratic S&P | Market S&P | Base Moody's |
| Capitalization (logarithm) | -0.617*** | 2.805* | -0.770*** | -0.445** |
|  | (0.178) | (1.475) | (0.177) | (0.813) |
| Squared |  | -0.426** |  |  |
|  |  | (0.182) |  |  |
| Return on assets | -0.063*** | -0.132*** |  | -0.065*** |
|  | (0.0178) | (0.032) |  | (0.019) |
| Squared |  | 0.0034*** |  |  |
|  |  | (0.0010) |  |  |
| EBITDA/Interest expenses | -0.011*** | -0.030*** | -0.016*** | -0.014* |
|  | (0.0040) | (0.010) | (0.0036) | (0.0079) |
| Squared |  | 0.00010** |  |  |
|  |  | (0.00005) |  |  |
| Long-term debt/Capital | 0.015*** | 0.021*** |  | 0.021*** |
|  | (0.0056) | (0.0068) |  | (0.0058) |
| Total debt/EBITDA | -0.059 | -0.215** |  |  |
|  |  | (0.092) |  |  |
| Cash flow/Sales |  |  |  | 0.019*** |
|  |  |  |  | (0.0072) |
| Proxy for current liquidity | 0.242* |  |  | 0.497*** |
|  | (0.142) |  |  | (0.154) |
| Volatility of value |  |  | 0.065*** |  |
|  |  |  | (0.012) |  |
| Share value/Cash flow |  |  | -0.025*** |  |
|  |  |  | (0.0086) |  |
| Telecommunication | -1.107** | -1.428*** | -0.430 | -1.638*** |
|  | (0.442) | (0.386) | (0.427) | (0.487) |
| Metal and mining | -1.514*** | -1.668*** | -1.702*** | -1.227** |
|  | (0.429) | (0.425) | (0.454) | (0.488) |
| Oil and gas | -1.884*** | -1.722*** | -1.733*** | -1.728*** |
|  | (0.491) | (0.392) | (0.427) | (0.481) |
| Consumer | -1.504*** | -1.893*** | -1.168** | -1.015* |
|  | (0.491) | (0.475) | (0.493) | (0.529) |
| Utilities | -2.795*** | -2.909*** | -1.804*** | -2.900*** |
|                          | (0.442)  | (0.441)  | (0.465)  | (0.513)  |
| Inflation level          | 0.463*** | 0.352*** | 0.567*** | 0.374*** |
|                          | (0.089)  | (0.074)  | (0.085)  | (0.086)  |
| GDP growth               | -0.171** | -0.197***| -0.262***| -0.029   |
|                          | (0.070)  | (0.069)  | (0.061)  | (0.086)  |
| Developed countries      | -0.714** | -1.170***|          | 0.334    |
|                          | (0.358)  | (0.362)  |          | (0.391)  |
| Pseudo-$R^2$             | 0.321    | 0.354    | 0.350    | 0.273    |
| Exact forecast Δ=0, %    | 39       | 37       | 43       | 42       |
| Not more than 1 class \|Δ\|≤1,% | 53 | 56       | 48       | 48       |

*, **, *** signify 10%-, 5%- and 1%-level of significance, respectively.

The signs for the indicators "Ratio of total debt to gross profit" and "Ratio of cash flow to volume of sales" require additional comment. The sign of the coefficients of the former can be explained by the high level of correlation with the ratio of long-term debt to capital (in percents). In a number of models, that indicator is meaningless, and its absence is not detrimental to the quality of the model. Similar explanations are possible for the indicator "Ratio of cash flow to sales volume." Total debt/EBITDA indicator is not significant or excluded from the models due to high correlation with Long-term debt/Capital ratio (see appendix 2).

The inclusion of macroeconomic indicators as well as the consideration of the factor of industry and country affiliation raised the quality of the base model to minimal acceptability. The influence of macroeconomic factors on the rating was expected: negative for inflation and positive for the GDP growth. This determines the level of stability of the external business environment.

Affiliation with developed countries in our research was not an obvious positive factor which was connected with the correlation present between this element and the macroenvironment indicator. Russian companies were not significantly distinguishable from companies in developing countries.

Affiliation with an industry has an influence on the rating. In particular, ratings of utilities companies, and those in the oil and gas sector are differentiated from heavy industry companies.

The introduction of quadratic dependences in a number of explanatory variables, i.e., capitalization, return on assets has improved the statistical characteristics of the model. The minimum point for return on assets and EBITDA to Interest expenses ratio lies outside the interval of significance of the variable. The sign for the coefficient is entirely defined by a linear member and the tendency is preserved. The variables used explain the ratings even without the use of the country rating.

The use of a stock market indicator (the market model) is an area which may be improved for models of publicly-traded companies (that is those with market quotations). Indicators of value volatility, level of systemic risk, ratio of share value to cash flow, as well as an indicator of market discipline in the country where the company was located (in the form of the corruption index) took place in our distribution.

Systemic risk was insignificant in practically all the models examined. Volatility of value negatively influenced the level of the rating because the market risks of the given asset grew when it took place. Growth of share value in relation to cash flow had a positive influence on the rating. The positive influence of capitalization, the ratio of gross earnings to interest expenses and return on assets was preserved, as was the influence of macroeconomic indicators. The index of corruption was insignificant in the models. This factor apparently has less influence on real production as compared with the administrative and financial spheres.

An analysis of the predictive power of the models was conducted by making a comparison of the true ratings of enterprises with their model values. Errors in forecast $\Delta$ as the difference between the forecasted and real ratings (in the numerical scale of classes), were used as a measurement. The accuracy of the forecast is on the level of 39-43%. The portion of forecasts that err by no more than one class is on the level of 90-92%. That is somewhat worse than for models elaborated for banks (Karminsky, Peresetsky, 2007).

That may be partly due to the insufficient volume of the sample, stratification by different industries and the use of a scale when there was a large quantity of ratings on the borders of class gradations. One more factor that is substantive for a comparison of forecast accuracy is the moment at which the sample is formed and the potential influence of the world financial crisis.

## 5.2. Models of corporate ratings on a mixed scale

As one of the areas for the improvement of the quality of the model, transition to a scale of gradations or a mixed scale may be considered. This would ensure the reasonable accuracy of the models. Relevant models are presented in Table 3.

Table 3

Models of ratings on scales of gradations and a mixed scale

|  | S&P | | S&P, market | | Moody's |
| --- | --- | --- | --- | --- | --- |
| Scale | Gradations | Mixed | Gradations | Mixed | Mixed |
| Volatility of value |  |  | 0.022*** | 0.068*** |  |
|  |  |  | (0.0060) | (0.011) |  |
| Share value/Cash flow |  |  | -0.015** | -0.26*** |  |
|  |  |  | (0.0075) | (0.0078) |  |
| Capitalization (logarithm) | -0.517*** | -0.509*** | -0.528*** | -0.588*** | -0.502*** |
|  | (0.151) | (0.153) | (0.154) | (0.158) | (0.158) |
| EBITDA/Interest expenses | -0.0062* | -0.0062* | -0.0089*** | -0.0086*** | -0.017** |
|  | (0.0034) | (0.0035) | (0.0033) | (0.0033) | (0.0070) |
| Return on assets | -0.035*** | -0.033** | -0.042*** | -0.041*** | -0.032** |
|  | (0.014) | (0.014) | (0.015) | (0.015) | (0.014) |
| Long-term debt/Capital | -0.012* | -0.012** |  |  | 0.0095* |
|  | (0.0045) | (0.0047) |  |  | (0.0049) |
| Inflation level | 0.379*** | 0.391*** | 0.443*** | 0.561*** | 0.345*** |
|  | (0.063) | (0.065) | (0.070) | (0.077) | (0.069) |
| GDP growth | -0.186*** | -0.184*** | -0.185*** | -0.252*** | -0.96 |
|  | (0.060) | (0.060) | (0.185) | (0.053) | (0.076) |
| Metal and mining |  |  | -0.456* | -0.835*** |  |
|  |  |  | (0.258) | (0.270) |  |
| Oil and gas | -0.619*** | -0.625*** | -0.866*** | -0.954*** | -0.413* |
|  | (0.197) | (0.198) | (0.212) | (0.215) | (0.228) |
| Utilities | -1.217*** | -1.223*** | -1.127*** | -0.973*** | -1.403*** |
| Developed countries | (0.0224) -0.636** (0.308) | (0.0225) -0.611** (0.310) | (0.234) | (0.238) | (0.243) 0.086 (0.355) |
| Pseudo-R$^2$ | 0.159 | 0.169 | 0.166 | 0.219 | 0.148 |
| Accuracy of forecast Δ = 0, | 34 | 31 | 35 | 39 | 33 |
| Error up to 1 gradation $|\Delta|$ ≤1,% | 52 | 57 | 51 | 50 | 57 |
| Error up to 2 gradations $|\Delta|$ ≤2,% | 14 | 12 | 13 | 10 | 9 |

*, **, *** signify 10%-, 5%- and 1%-level of significance, respectively.

For models using market indicators, as before, volatility of a company's share value exerted a negative influence. The influence of the ratio of share value to cash flow was positive. Among the balance indicators, size of the company (capitalization) retained its positive influence. Profit to interest expenses ratio and return on assets did too. The influence of the settlement of long-term debt to capital ratio was also positive. This was due to the high negative correlation of this variable with the three previous ones.

Level of inflation was the definitive macroeconomic variable. The relevant coefficients were significant in all models and had negative influence. GDP growth had positive influence, but its presence in the models may have varied because of its high correlation with the inflation indicator. The level of corruption was not included in the model. This was due in part to the presence of affiliation with the developing countries dummy among the explanatory variables.

The positive influence on the ratings of companies affiliated with the oil, gas and utilities industries could be noted. Moreover, an analysis of average deviations showed that this tendency was stable.

Accuracy of forecasts with an error of 1 gradation was about 90% and with an error of no more than 2 gradations was higher than 99%, which was better than the error for models in the scale of classes. Distribution of errors for the market model on a mixed scale is shown in the bar chart (Fig. 4). For that model, errors of the first type on the level of 1 gradation did not exceed 4-5%.

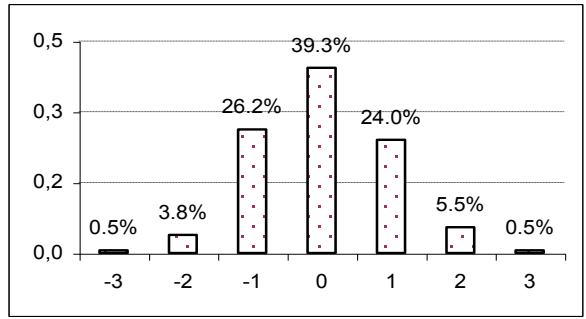

Fig. 4. Bar chart of errors for the market model

## 5.3. Rating models for banks

Modeling bank financial strength rating (BFSR) and bank deposits rating were presented in grades and had nearly the same quality of models forecast. The following conclusions may be drawn from an analysis of these (Peresetsky and Karminsky, 2008).

Banks located in developing countries have a lower BFSR than banks in developed countries. The negative significance of the explanatory dummy of affiliation with developing markets, which is quite logical, should also be brought to attention. The difference of evaluations of Russian, Kazak and Ukrainian banks in comparison with banks of developing market should also be noted (for reasons that include the level of corruption and political risk). This indicates the potential for banks' ratings growth in CIS countries.

The volume of assets has a positive effect. Parameters reflecting efficiency (ratio of personnel expenses to operating income), the quality of assets (ratio of overdue debt to all debt) and ratio of client funds to own capital are significant. Growth of personnel expenses, bad debts and financial leveraging has a negative influence on the BFSR.

The cost of liabilities has negative influence what means that banks which payed more to obtain funds have a worse BFSR indicator. The cost of resources determines the level of stability and efficiency of banks' activities.

# 6. Comparative analysis of the corporate ratings of the two agencies

We made a statistical comparison of the ratings of the S&P and Moody's agencies. We used a subsample containing observations of companies that had ratings from both agencies. This sample amounted to 178 companies.

Three measures of difference were used for the comparison:

$\Delta$ - the difference between S&P and Moody's ratings

FDS = $|\Delta|$

SPLIT - a binary function that takes the value of 0 when the ratings coincide and 1 otherwise.

For each measure, we elaborated econometric models to determine the factors that significantly influenced the distribution of opinions in the agencies as expressed in their ratings. The results of the comparison were presented in Table 4.

The following conclusions may be made from an analysis of the table:

1. The most substantial difference was the rating of companies from developing countries. It was expressed either directly through the appropriate dummy of affiliation with developed countries or indirectly as the influence of corruption.

2. Among the most significant and positive factors influencing the ratings of Moody's agency, return on assets may be noted. For the other agency, factors such as instant liquidity, share of fixed assets in total assets, level of inflation and corruption were more significant.

Table 4

Comparison of the ratings of the S&P and Moody's agencies

|  | Model number | | | | |
|---|---|---|---|---|---|
|  | Difference Δ | | Difference module ABS | | SPLIT distinction |
|  | 1d | 2d | 1a | 2a | 1s |
| Return on assets | 0.028** (0.011) | 0.022* (0.011) | | | |
| Instant liquidity | -0.462*** (0.140) | -0.507*** (0.150) | | | |
| Fixed assets/Assets | | -0.924** (0.383) | | | |
| Share value/Cash flow | | -0.0098 (0.0073) | | | |
| Value volatility | | | -0.007* (0.0036) | -0.006 (0.0036) | -0.0038* (0.0023) |
| Inflation level | -0.221*** (0.056) | -0.156*** (0.060) | | | |
| Corruption index | -0.303*** (0.060) | -0.305*** (0.058) | -0.81 (0.053) | | |
| Consumer sector | -0.857*** (0.284) | -1.084*** (0.288) | -0.169 (0.216) | | |
| Developed countries | | | -0.572** (0.237) | -0.838*** (0.163) | -0.309*** (0.086) |
| Russia | | | -0.649*** (0.243) | -0.098 (0.341) | |
| Other insignificant | | | | + | |
|  | 0.182 | 0.217 | 0.127 | 0.140 | 0.061 |

3. No substantial difference in the ratings of Russian companies was uncovered. The positive influence of the dummy of affiliation with Russia indicated the presence of large differences in ratings both on the positive and negative sides (model 1a).

4. Growth of the volatility of companies' share value creates multidirectional differences, although not at a very high level of significance – 10% (models 1a and 1s). This indirectly confirmed the previous conclusion.

5. The S&P agency takes a more critical position toward companies from the consumer sector (models 1d and 2d). On average, the divergence between the agencies' ratings, expressed as their difference Δ, was 0.26 gradation for our sample and is characterized by the bar chart in Fig. 5.

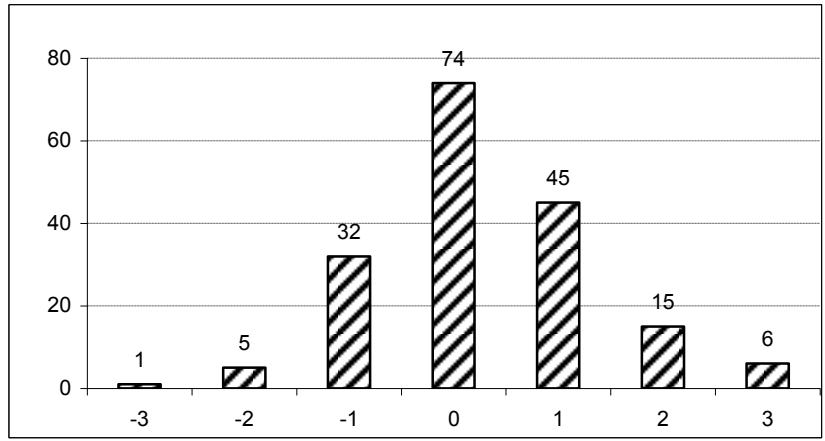

Fig. 5. Sample distribution of the difference between the ratings of the S&P and Moody's agencies

# Conclusion

Remote probability evaluations based on econometric models should be an integral part of internal rating systems which determine the potential practical significance of such models, especially in developing countries. In this work, rating models of industrial enterprises and of the financial stability of corporations and banks were elaborated based on multiple choice models. Financial indicators of corporations and banks, dummies of regional affiliation of banks and years were employed as explanatory variables.

It was demonstrated that:
1. When other conditions are equal, industrial companies and banks in developing countries receive lower ratings in comparison with banks in developed countries. Factors which determine lower ratings for emerging countries require additional analysis as well as factors determining higher market volatility in such countries.
2. Ratings depend on affiliation with an industry. The upper ratings have utility companies, also as oil&gas, metal and mining and consumer industries.
3. A set of explanatory financial indicators is sufficient and easily inter preted to rating models. The degree of influence of country affiliation, re turn on assets, instant liquidity, inflation level and corruption are prominent factors that differentiate between the approaches of the two agencies.
4. The predictive power of models of corporate ratings and bank financial strength ratings is somewhat better than deposit rating models. The deviations present are mostly explainable by quality factors.

# Appendix 1

**Distribution of ratings of developing countries**

Distribution of CEE countries ratings

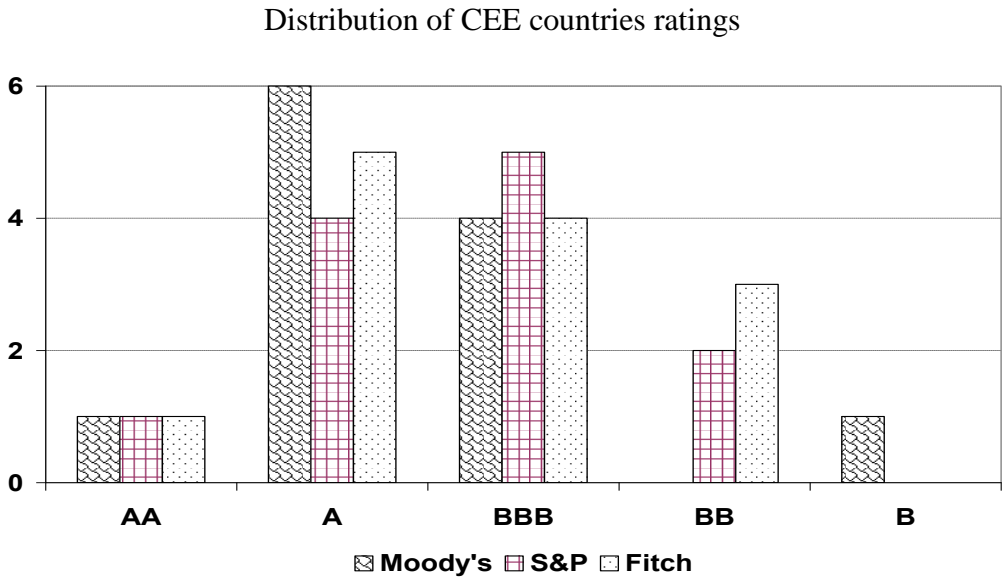

Distribution of banks developing countries ratings

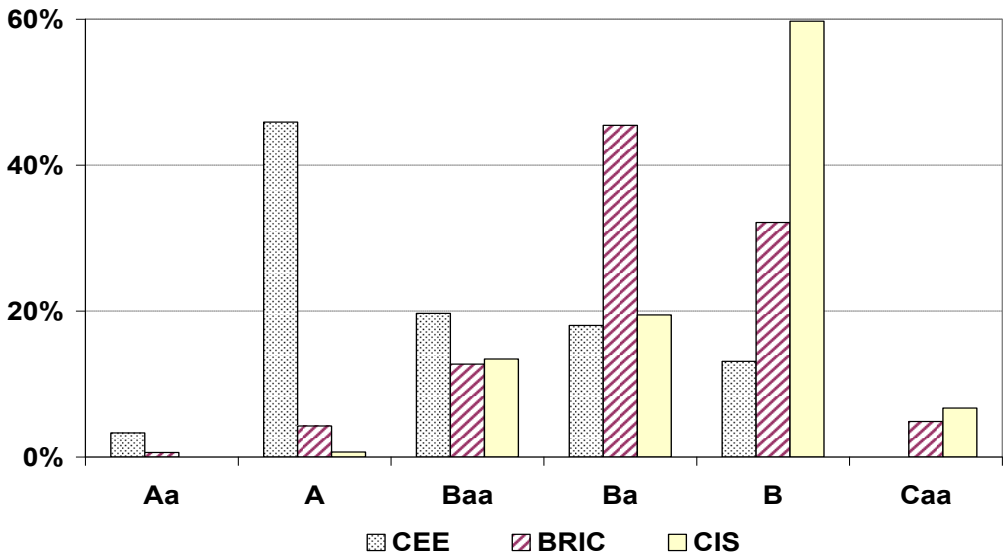

# Appendix 2

**Descriptive statistics and sample correlation matrix for industrial enterprises**

**Descriptive statistics**

|  | Return on assets | EBITDA/Interest expenses | Total debt/ EBITDA | Cash flow/Sales | Operating margin | Proxy for current liquidity | Capitalization (logarithm) | Long-term debt/ Capital |
|---|---|---|---|---|---|---|---|---|
| Mean | 8.14 | 18.64 | 2.03 | 22.89 | 19.10 | 1.29 | 4.16 | 34.31 |
| Median | 7.18 | 10.81 | 1.50 | 18.50 | 16.82 | 1.10 | 4.19 | 30.62 |
| Maximum | 40.56 | 241.77 | 9.63 | 72.87 | 59.53 | 4.06 | 5.67 | 149.16 |
| Minimum | -12.82 | 1.240 | 0.030 | 0.38 | -6.10 | 0.17 | 2.33 | 0.01 |
| Standard deviation | 6.39 | 25.84 | 1.67 | 15.05 | 12.40 | 0.71 | 0.55 | 20.18 |

**Correlation matrix**

|  | Return on assets | EBITDA/Interest expenses | Total debt/EBITDA | Cash flow/Sales | Operating margin | Proxy for current liquidity | Capitalization (logarithm) | Long-term debt/ Capital |
|---|---|---|---|---|---|---|---|---|
| Return on assets | 1.00 | 0.422 | -0.564 | 0.311 | 0.563 | 0.259 | 0.290 | -0.333 |
| EBITDA/Interest expenses | 0.422 | 1.00 | -0.426 | 0.128 | 0.211 | 0.183 | 0.176 | -0.445 |
| Total debt/EBITDA | -0.564 | -0.426 | 1.00 | -0.253 | -0.357 | -0.199 | -0.235 | 0.688 |
| Cash flow/Sales | 0.311 | 0.128 | -0.253 | 1.00 | 0.781 | -0.174 | 0.016 | -0.052 |
| Operating margin | 0.564 | 0.211 | -0.357 | 0.780 | 1.00 | 0.075 | 0.073 | -0.144 |
| Proxy for current liquidity | 0.259 | 0.183 | -0.198 | -0.174 | 0.075 | 1.00 | 0.015 | -0.212 |
| Capitalization (logarithm) | 0.290 | 0.176 | -0.235 | 0.016 | 0.073 | 0.015 | 1.00 | -0.268 |
| Long-term debt/Capital | -0.333 | -0.445 | 0.688 | -0.052 | -0.144 | -0.212 | -0.269 | 1.00 |

Descriptive statistics

| | Return on assets | EBITDA/Interest expenses | Total debt/ EBITDA | Cash flow/Sales | Operating margin | Proxy for current liquidity | Capitalization (logarithm) | Long-term debt/ Capital |
|---|---|---|---|---|---|---|---|---|
| Mean | 8.14 | 18.64 | 2.03 | 22.89 | 19.10 | 1.29 | 4.16 | 34.31 |
| Median | 7.18 | 10.81 | 1.50 | 18.50 | 16.82 | 1.10 | 4.19 | 30.62 |
| Maximum | 40.56 | 241.77 | 9.63 | 72.87 | 59.53 | 4.06 | 5.67 | 149.16 |
| Minimum | −12.82 | 1.24 | 0.03 | 0.38 | −6.10 | 0.17 | 2.33 | 0.01 |
| Standard deviation | 6.39 | 25.84 | 1.67 | 15.05 | 12.40 | 0.71 | 0.55 | 20.18 |